\documentclass[12pt]{iopart}

\usepackage{graphicx}

\usepackage{array}

\def\be{\begin{equation}}
\def\ee{\end{equation}}
\def\bea{\begin{eqnarray}}
\def\eea{\end{eqnarray}}
\def\ba{\begin{array}}
\def\ea{\end{array}}
\def\bdm{\begin{displaymath}}
\def\edm{\end{displaymath}}

\begin{document}

\title[Thermal Hall Conductivity of a Nodal Chiral Superconductor  ]
{Low Temperature Thermal Hall Conductivity of a Nodal Chiral Superconductor}

\author{Sungkit Yip}

\address{Institute of Physics and Institute of Atomic and Molecular Sciences\\
 Academia Sinica \\
128 Academia Road, Sec 2, Nankang,
Taipei 115, Taiwan.}
\ead{yip@phys.sinica.edu.tw}
\begin{abstract}

Motivated by Sr$_2$RuO$_4$, we consider a chiral
superconductor where the gap is strongly suppressed
along certain momentum directions.
We evaluate the thermal Hall conductivity
in the gapless regime, {\it i.e.}, at temperature
small compared with the impurity band width $\gamma$,
taking the simplest model of isotropic impurity scattering.
We find that, under favorable circumstances,
this thermal Hall conductivity can be quite significant,
and is smaller than the diagonal component
(the universal thermal conductivity) only by
a factor of $1/ \ln ( 2 \Delta_M/\gamma)$, where
$\Delta_M$ is the maximum gap.

\noindent 

\end{abstract}

\maketitle

\section{Introduction}\label{sec:intro}

Impurities play a dual role in transport properties of unconventional
superconductors.   They scatter the carriers responsible for the transport.
This effect decreases the transport coefficients.  On the other hand,
impurities are pair-breaking and so generate new excitations which
contribute to the transport.  Under suitable circumstances, these
two effects exactly cancel with each other and give rise to
universal transport coefficients, independent of
the concentration or other properties specifying the
impurities.   This was first discussed theoretically
by P. A. Lee \cite{Lee93} for the (low temperature
finite frequency) electrical conductivity
in high temperature cuprate superconductors with $d_{x^2 -y^2}$ order parameter.
Subsequently, the author
and his collaborators extended this result to {\it thermal conductivity}
(as well as other quantities such as ultrasonic attenuation coefficients)
and also for other nodal superconductors \cite{Graf96,Graf00}.
These universal transport coefficients apply
at the low temperature, low frequency
(temperature and frequency small compared with impurity band
width $\gamma$) regime, which for definiteness will be referred
to as {\em gapless} (in analogy with the corresponding regime
for conventional superconductors with magnetic impurities).
This {\em universal thermal conductivity} has played an
extremely useful role in deducing the nodal structures of unconventional
superconductors (see, e.g. \cite{Taillefer} for a review).
The success of this is partly due to the absence of impurity vertex corrections
of these transport coefficients under many circumstances \cite{Graf96},
and even when these corrections are present, may give rise only to
a relatively small contribution \cite{Durst}.

Sr$_2$RuO$_4$, a superconductor with similar crystal structure
to the cuprate superconductors with also basically only two-dimensional
dispersing bands,
has long been proposed to be a superconductor with a
chiral order parameter breaking time-reversal and reflection
symmetries (see \cite{Maeno12} for a review), though
the actual form of the order parameter is far from certain
and highly controversial, with the situation further complicated
by the presence of three conduction bands at the Fermi level.
Initially it was proposed that the order parameter has
the momentum dependence $~ p_x + i p_y$, and is fully gapped.
However, numerous experiments such as specific heat
\cite{Nishizaki00,Deguchi04}, have shown that this superconductor
has low lying excitations.  Thus the order parameter
possesses nodes, or at least momentum directions
where there are strong suppressions of
the gap, which we shall referred to as ``near-nodes".
In particular, thermal conductivity
at the zero temperature limit is as expected from the theory
of superconductor with line nodes \cite{Suzuki02,Suderow98,Graf00b}
(point nodes if we regard the system as two-dimensional).
These line nodes or near-nodes have been attributed
to the momentum dependence of the spin fluctuation responsible
for the pairing, proximity to the Brillouin zone boundary,
spin-orbit couplings, or combinations there-of
\cite{Graf00b,Nomura02,Yanase03,Wang13,Scaffidi15}.  We note
here that these nodes or near-nodes are not necessarily
associated with sign change of the order parameter:
indeed no such sign change is needed at least for the
scenario in \cite{Nomura02} where the (near-)node occurs for
the band labeled by $\gamma$
(not to be confused with the impurity band width)
at momenta near the zone boundaries
(no bands are believed to actually cross or touch the zone boundaries).

For a chiral superconductor, off-diagonal elements in
transport coefficients such as Hall conductivity are symmetry allowed.
If the electrical Hall conductivity of a system
is finite, then an electric field in the say,
x-direction can produce an electric current along the y-direction,
or vice versa.  Similarly, one can also have Hall thermal conductivity
where a temperature gradient along x produces an energy current
along y in addition to one along x.
However the existence of such Hall transport coefficients turns out to be tricky.
(Consider the simplest example of a clean superconductor with a quadratic
normal state dispersion.  A uniform electric field can only
accelerate the system as a whole and hence the electric Hall conductivity
must be identically zero \cite{Yip92}).
Thermal Hall conductivity of chiral superconducting states has in
fact been examined theoretically long time ago by Arfi {\it et al} \cite{Arfi}.
These authors confined themselves to the case where quasiparticles are well-defined
(thus at temperatures large compared with the impurity band-width
mentioned above).  Here once again impurities play an important role.
 Considering isotropic scatterers,
these authors have shown that the Hall thermal conductivity
arises entirely from the  ``in"-scattering term
of the collision integral.
In addition to breaking
time-reversal and reflection symmetries, one must break particle-hole symmetry
as well.   The impurities once again provide
the necessary mechanism, if
the scattering phase shift $\delta$ of the impurities is not a
multiple of $\pi/2$.   For chiral superconductors,
these authors only examined
a three-dimensional superconductor with order parameter
$~ (p_x + i p_y)$.  This superconductor is fully gapped
near the basal plane, with two nodes in the north and south
poles which however contribute little to transport
at low temperatures.  Not surprising they obtain a Hall thermal
conductivity which is very small at low temperatures.

Returning to Sr$_2$RuO$_4$, a (basically) two-dimensional
superconductor with line nodes and (probably) an order parameter with
chiral symmetry, it is therefore highly interesting to
evaluate theoretically the expected thermal Hall conductivity.
Furthermore, given the observed universal thermal conductivity
at low temperatures,  it is important to examine this
thermal Hall conductivity in this
corresponding gapless regime, where only
the excitations near the nodes contribute
to the thermal transport.   We report such a calculation here,
employing the quasiclassical method as in \cite{Graf96}.
This thermal Hall conductivity
arises from the corrections to the impurity self-energies
(corresponding to vertex corrections in the response function language
and in-scattering in the Boltzmann kinetic equation approach).
While it vanishes for very weak ($\delta \sim 0 $ or integral multiples of $\pi$)
or very strong scatterers ($\delta \sim \pi/2$ modulo $\pi$),
under favorable circumstances it is
smaller than the universal diagonal thermal conductivity
only by the factor $ \sim 1/ \ln (\Delta_M/ \gamma)$, where
$\Delta_M$ is the maximum gap, and $\gamma$ the impurity band
width.  Since
this is only logarithmically small, this quantity may
be experimentally measurable.  The existence of this thermal Hall conductivity
would be a strong indication that the superconductor is
indeed a chiral superconductor.

Sr$_2$RuO$_4$ is a multi-band superconductor.
  The normal
state has three bands $\alpha,\beta,\gamma$ (not to be
confused with the impurity scale mentioned above) at
the Fermi level. There are
some controversies on which band dominates the thermodynamic
or transport properties, and how much
each band contributes to each quantity
 (compare e.g. \cite{Nomura02,Yanase03,Wang13,Raghu}).
For simplicity, we shall present the calculation for
a single band superconductor, assuming that it possesses
a chiral but nodal (or nearly nodal) order parameter.   The observed universal
thermal conductivity can be explained if there are line
nodes (or nearly nodes) on each of the three bands, or if say the $\gamma$-band
dominates the transport with this band alone having
a line node.    For either case however our calculation
below gives a semi-quantitative prediction of the
thermal Hall conductivity, assuming that each band
contributes to this quantity in parallel and hence
the total is simply given by the sum of the contribution
from each band.    If a particular band
is fully gapped and has a gap magnitude large compared
with the impurity scale $\gamma$, then that band
would have no contribution to the thermal transport
at low temperatures.

  The rest of this paper is organized as follows.  In \sref{sec:cal}
  we present first the general formulation, before specifying
  to a particular form of the gap.  We then estimate the
  thermal Hall conductivity.  \Sref{sec:sum} contains a discussion
  and summary. \Sref{sec:AppA} evaluates also the vertex corrections
  to the diagonal thermal conductivity, while \sref{sec:AppB}
  contains discussions for more general forms of the gap.

\section{Thermal Hall conductivity}\label{sec:cal}

Our calculation is an extension of \cite{Graf96},
and we would employ similar notations except otherwise
stated (see also \cite{Xu95}).
We shall assume that the order parameter is
given by $\hat z ( \Delta_x + i \Delta_y)$,
where $\Delta_{x,y}$ transforming as a two-dimensional
representation under the tetragonal symmetry of
the crystal.   Here we have taken a simple momentum
independent spin structure of the order parameter
$\hat z$ implying that we have only up-down pairing.
Some recent models \cite{Wang13,Scaffidi15}
suggest that this is too simplistic but relaxing
this assumption is expected only to give rise to some slightly
different numerical factors for our main results:
see below.  To simply our presentation we shall
also take a cylindrical Fermi surface: generalizations
to angular dependent density of states and Fermi velocities
are straight-forward but would make notations rather
clumsy (it is here worth remembering that the low temperature
transport must be dominated by contributions near
the nodes).   For $\Delta_{x,y}$ above, we take the model

\bea
\Delta_x (\hat p) &=& \Delta_M \ \cos \phi_{\bf \hat p} \ \eta_x(\phi_{\bf \hat p})
\nonumber \\
\Delta_y (\hat p) &=& \Delta_M \ \sin \phi_{\bf \hat p} \ \eta_y(\phi_{\bf \hat p})
\label{Dx}
\eea
where $\phi_{\bf \hat p}$ is the azimuthal angle of the momentum direction
$\bf \hat p$.  For simplicity we shall often leave out
the subscript $\bf \hat p$ for $\phi$.
Without the $\eta_{x,y}$ factors, the magnitude
of the gap at any $\bf \hat p$ is given by
$(|\Delta_x|^2 + |\Delta_y|^2)^{1/2} = \Delta_M$, our maximal
gap.  To provide a model consistent with universal thermal
conductivity, we need line or near-line nodes.
We do this by the factors $\eta_{x,y}$.
For the moment, we do not need their specific form,
but only need to note that they are such that
they do not change the symmetry properties of
the pair $ (\Delta_x, \Delta_y)$ in the sense
that they must still transform as the two components
of a two-dimensional representation
under tetragonal symmetry.  In particular, $ (\Delta_x, \Delta_y)$ transform
in the same manner as $(v_{fx}, v_{fy})$, the Fermi velocity components
along $x$ and $y$.

The quasiclassical Green's function $\check g$ obeys
\be
[\epsilon \tau_3 - \hat \Delta - \check \sigma_{\rm imp}, \check g]
+ i {\bf v_f} \cdot {\bf \nabla} \check g = 0
\label{qc}
\ee
\be
\check g^2 =  - \pi^2 \check 1
\label{norm}
\ee
where $\epsilon$ is the energy,
$\check \sigma_{\rm imp}$ the impurity self-energy,
$v_f$ the Fermi velocity,
${\bf \nabla}$ the spatial gradient.
The check symbol ( $\check{}$ ) denotes matrix
in Keldysh (R, K, A) space, and $ [ \cdots, \cdots]$ the commutator.
 $\hat \Delta$, the off-diagonal fields for superconducting pairing,
  is diagonal in the Keldysh space
but a matrix in particle-hole and spin space.
$\hat \Delta = \sigma_1 \left( (\Delta_x + i \Delta_y) \tau_+
 - ( \Delta_x - i \Delta_y)  \tau_- \right) $ where
 $\tau_{3,+,-}$ ($\sigma_1$) are the Pauli matrices in particle-hole
 (spin) space.
  Note $\hat \Delta^2 = - | \Delta (\phi)|^2$ $= -(\Delta_x^2 + \Delta_y^2)$.

 Assuming isotropic scattering, the impurity
self-energy $ \check{\sigma}_{\rm imp}$ is given by
\be
\check{\sigma}_{\rm imp} = \ n_{\rm imp} \check {t}
\label{simp}
\ee
where
\be
\check t \equiv  u
\left( 1 -  N_f u \langle \check{g} \rangle \right)^{-1}
\label{t}
\ee
Here $n_{\rm imp}$ is the impurity density,
$u$ the impurity potential, $N_f$ the density of states
per spin,  and the angular brackets $ \langle  \cdots \rangle $
 denote angular average of the quantity within the brackets.
 Defining $\Gamma_u = n_{\rm imp} / \pi N_f$ and
 $ \cot \delta = - 1/ ( \pi N_f u)$,
 \eref{simp} can be rewritten as
 \be
 \check{\sigma}_{\rm imp} = -  \ \Gamma_u
 \left( {\rm cot} \delta + \langle \frac{\check{g}}{\pi} \rangle \right)^{-1}
 \label{simp2}
 \ee
 (Our expression relating $\delta$ and $u$ has a different
 sign from \cite{Xu95}.  With the present definition,
 the retarded component of \eref{t} in the normal
 state, with $\hat g^R = - i \pi \tau_3$, reads
 $\hat t^R = - \left( \pi N_f ( {\rm cot} \delta - i \tau_3)\right)^{-1}$
 so that the particle component is
 $  - \frac{ 1 } {\pi N_f} \sin \delta \ e^{ i \delta}$,
 the usual convention. )

First we consider the uniform equilibrium case,
with quantities distinguished by the subscripts $0$.
  The retarded components of
\eref{simp} and \eref{qc} imply that in equilibrium
we have, using $ \langle \hat \Delta \rangle = 0$,
\be
\hat g_0^R (\hat {\bf p}, \epsilon) =
 -\pi \frac{ \tilde{\epsilon}^R \tau_3 - \hat {\Delta}}{D^R}
 \label{g0R}
 \ee
 where $D^R \equiv \sqrt{ | \Delta (\phi)|^2 - (\tilde {\epsilon}^R)^2}$.
 (This quantity was written as $ - \pi C^R$ in \cite{Graf96}.)
  We shall often write
  $\hat M^R \equiv \tilde{\epsilon}^R \tau_3 - \hat {\Delta}$,
  with $(\hat M^R)^2 = - (D^{R})^2$.  Since we are
  interested in the low temperature limit, we can
  focus on the
 the limit $\epsilon \to 0$ for our Green's functions.
 In this case we have $\tilde {\epsilon}^R \to i \gamma$
 with $\gamma>0$, which arises from the $\tau_3$ component of
 $\sigma^R_{0, \rm imp}$, obeying the self-consistent equation
 \be
 \gamma = \Gamma_u
 \frac{ \tilde \gamma }
{ {\rm cot}^2 \delta
 + \tilde \gamma^2}
 \label{gamma}
 \ee
where we have defined the dimensionless quantity
\be
\tilde \gamma \equiv \langle
\frac{ \gamma} {(|\Delta (\phi)|^2 + \gamma^2)^{1/2}} \rangle
\label{tgamma}
\ee
  The advanced components
 are given by similar equations (except, e.g.
 $\tilde \epsilon^A \to - i \gamma$.)
 Rigorously speaking, $\sigma^{R,A}_{0, \rm imp}$ both also contain a
 part which is proportional to $\tau_0$.  For
 $\epsilon \to 0$ they are both given by
$ - \Gamma_u  {\rm cot} \delta /
\left( {\rm cot}^2 \delta + \tilde \gamma^2
\right) $
 and is thus independent of energy.  This
 quantity drops out of all our equations below
 (as physically expected since it just represents an overall
 shift in the energy).
 Note that, in this limit, we also have
 $D^R = D^A \to D \equiv (|\Delta (\phi)|^2 + \gamma^2)^{1/2}$.
 Generally in equilibrium
 $\hat g_0^K = (\hat g_0^R - \hat g_0^A) h(\epsilon, T)$
 where $h(\epsilon, T) = \tanh (\epsilon/2T)$ where $T$ is
 the temperature.

 In the presence of a temperature gradient,
 $T$ becomes position dependent.
 $\Delta_{x,y}$ etc are implicitly temperature ($T$) dependent
 (but not $n_{\rm imp}$ nor $u$ characterizing the impurities).
 The quasiclassical Greens function
 to zeroth order in the gradient is still given by
 expressions given above though the temperature variable there
 is position dependent.   To obtain the first
 order correction to the Green's functions,
 we perform a corresponding expansion of the
 quasiclassical equation \eref{qc}.
  $\hat g_1^R (\hat {\bf p}, \epsilon)$,  the
 first order correction to the retarded quasi-classical Green's
 function obeys
 \be
 [\tilde \epsilon^R \tau_3 - \hat \Delta, \hat g_1^R]
 - [ \hat \sigma_{1, \rm imp} ^R + \hat \Delta_1, \hat g_0^R]
 + i {\bf v_f} \cdot {\bf \nabla} \hat g_0^R = 0
 \label{qcR}
 \ee
 Here $\sigma_{1, \rm imp}^R$ and $\hat \Delta_1$
 are the first order correction to the impurity and
 pairing self-energies.  These two quantities need
 to be found self-consistently.  We shall see that
 $\hat \Delta_1 = 0$ after we evaluate the Kelydsh
 component of the quasiclassical Green's function below
 (since $\hat g_1^K$ is odd in $\epsilon$).
 \Eref{qcR} can be solved \cite{Xu95} by exploiting
 the fact that $\hat M^R$ and $\hat g_1^R$ anti-commute
 (since $\hat g_0^R$ and $\hat g_1^R$ anti-commute, thanks
 to the normalization condition
 $\hat g^2 = - \pi^2$).  We obtain
 \be
 \hat g_1^R = \frac{\hat M^R}{ 2 D^2}
 \left(  i {\bf v_f} \cdot {\bf \nabla} \hat g_0^R
   - [\hat \sigma^R_{1, \rm imp}, \hat g_0^R] \right)
   \label{g1R}
 \ee
 Together with the retarded component of \eref{simp2},
 one can obtain a self-consistent equation for
 $\sigma^R_{1,\rm imp}$ and hence $\hat g_1^R$.
 In the first term, the gradient arises only through
 the dependence of $\Delta_M$
 (and hence indirectly $\tilde \epsilon^R$) on the temperature $T$.
 An explicit evaluation shows that the angular average of
 this term vanishes.  Hence it is consistent to set
 $\sigma^R_{1,\rm imp} = 0$.
 However, we shall not need an explicit expression
 for $\hat g_1^R$ below, since the trace
 ${\rm Tr} (g_1^R)$ vanishes.  (The second term in
 \eref{g1R}, even when $\hat \sigma_{1, \rm imp}^R$ were finite,
  has no trace.  The first term
 also has no trace since ${\rm Tr} (\hat g_0^R \nabla \hat g_0^R)
 = \frac{1}{2} {\rm Tr}  \nabla (\hat g_0^R \hat g_0^R) = 0$,
 as noted already in \cite{Graf96}.)
 $\hat g_1^A$ is given by an equation similar to \eref{g1R}
 except $R \to A$, and we also have ${\rm Tr} (g_1^A) = 0$.

 The Kelydsh component of the \eref{qc}, to first
 order in gradient, reads
 \bea
 (\hat M^R \hat g_1^K - \hat g_1^K \hat M^A)
  - (\sigma_{0, \rm imp}^K \hat g_1^A - \hat g_1^R \hat \sigma^K_{0, \rm imp})
  \nonumber \\
 - (\sigma_{1, \rm imp}^R \hat g_0^K - \hat g_0^K \hat \sigma^A_{1, \rm imp})
 - (\sigma_{1, \rm imp}^K \hat g_0^A - \hat g_0^R \hat \sigma^K_{1, \rm imp})
    + i {\bf v_f} \cdot {\bf \nabla} g_0^K = 0
 \label{qcK}
 \eea
 (In writing \eref{qcK}, we have already used the fact that
 at $\epsilon \to 0$, the $\tau_0$ parts of $\sigma_{0, \rm imp}^{R,A}$
 cancel each other).
 To simply this equation, it is convenient to
 write
 $\hat g_1^K = (\hat g_1^R - \hat g_1^A) h(\epsilon, T)
  + \hat g_{1a}^K$ which defines $\hat g_{1a}^K$.
$(\hat g_1^R - \hat g_1^A) h(\epsilon, T)$
and
$\hat g_{1a}^K$ are referred to as the ``regular" and
``anomalous" parts
of $\hat g_1^K$, in analogy to what appear in calculation of
response functions in diagrammatic methods.
We also write a similar equation for the impurity
self-energies, thus
$\hat \sigma_{1, \rm imp}^K =
 (\hat \sigma_{1, \rm imp}^R - \hat \sigma_{1, \rm imp}^A) h(\epsilon, T)
  + \hat \sigma_{1a, \rm imp}^K$.
  With the help of \eref{qcR} and the
analogous formula for $\hat g_1^A$ we obtain
\be
\hat M^R \hat g_{1a}^K - \hat g_{1a}^K \hat M^A
- (\sigma_{1a, \rm imp}^K \hat g_0^A - \hat g_0^R \hat \sigma^K_{1a, \rm imp})
    + i (\hat g_0^R - \hat g_0^A) {\bf v_f} \cdot {\bf \nabla} h(\epsilon,T) = 0
    \label{qcKa}
\ee
Expanding \eref{simp2} to first order and reading off the $K$ component,
we can verify that, were it not for $\hat g_{1a}^K$,
$\sigma_{1a, \rm imp}^K$ would vanish, that is,
the contribution to  $\sigma_{1a, \rm imp}^K$
from the ``regular part" of $\hat g_1^K$ and $\hat g_1^{R,A}$
cancel exactly.  Therefore we are left with
\be
\sigma_{1a, \rm imp}^K =
\Gamma_u \left( {\rm cot} \delta + \langle \frac{\hat g_0^R}{\pi} \rangle
\right)^{-1}
   \  \langle \frac{\hat g_{1a}^K}{\pi} \rangle \
    \left( {\rm cot} \delta + \langle \frac{\hat g_0^A}{\pi} \rangle \right)^{-1}
\ee
and in the low energy limit,
 \be
\sigma_{1a, \rm imp}^K =
\frac{ \Gamma_u} { \mathcal{C}}
\left( {\rm cot} \delta + i \tilde \gamma \tau_3 \right)
   \  \langle \frac{\hat g_{1a}^K}{\pi} \rangle \
 \left( {\rm cot} \delta - i \tilde \gamma \tau_3 \right)
 \label{s1aK}
\ee
and
\be
\mathcal{C} \equiv ({\rm cot}^2 \delta + \tilde \gamma^2 )^2
\label{C}
\ee
Using the Keldysh component of \eref{norm} and the
definition of $\hat g_{1a}^K$, we get
$\hat g_0^R  \hat g_{1a}^K + \hat g_{1a}^K \hat g_0^A = 0$,
which allows us \cite{Graf96} to eliminate
$\hat g_{1a}^K \hat M^A$ in favor of $\hat M^R \hat g_{1a}^K$.
$\hat g_{1a}^K$ can then be found.  It is convenient to
split it into two parts:
\be
\hat g_{1a}^K = \hat g_{1a}^{K, \rm ns} + \hat g_{1a}^{K, V}
\label{g1aK}
\ee
where
\be
\hat g_{1a}^{K, \rm ns} =
\frac{M^R}{D^R (D^R + D^A)}
\{ (\hat g_0^R - \hat g_0^A)
 ( - \frac{\epsilon}{2 T^2} {\rm sech}^2 \frac{\epsilon}{2T} )
 ( i {\bf v_f} \cdot {\bf \nabla} T )  \}
\label{g1aKns}
\ee
and
\be
\hat g_{1a}^{K, \rm V} = -
\frac{M^R}{D^R (D^R + D^A)}
\{
 (\sigma_{1a, \rm imp}^K \hat g_0^A - \hat g_0^R \hat \sigma^K_{1a, \rm imp})
  \}
\label{g1aKV}
\ee
The first term $\hat g_{1a}^{K, \rm ns}$ is the answer we would
get if the correction to the impurity self-energy $\sigma_{1a, \rm imp}^K$
was not included.  We denote it by ``${\rm ns}$" and refer it as
the ``non-self-consistent" contribution.  The contribution
$\hat g_{1a}^{K, \rm V}$ is directly proportional
to $\sigma_{1a, \rm imp}^K$ and is referred to as the ``vertex correction"
in analogy to the corresponding quantity
in diagrammatic response function calculations.

$\hat g_{1a}^{K, \rm ns}$ can be
directly evaluated to be, in the $\epsilon \to 0$ limit,
\be
\frac{ \hat g_{1a}^{K, \rm ns}}{\pi} =
 \frac{ i \gamma^2 + \gamma \tau_3 \hat \Delta}
{(|\Delta (\phi)|^2 + \gamma^2)^{3/2}}
       (  \frac{\epsilon}{2 T^2} {\rm sech}^2 \frac{\epsilon}{2T} )
  {\bf v_f} \cdot ( - {\bf \nabla} T )
 \ee
 with the corresponding angular average
 \be
 \langle \frac{ \hat g_{1a}^{K, \rm ns}}{\pi} \rangle =
 - \gamma (  \frac{\epsilon}{2 T^2} {\rm sech}^2 \frac{\epsilon}{2T} )
   \tau_3 \hat \Lambda
   \label{g1aKnsav}
   \ee
   where the dimensionless matrix $\hat \Lambda$ is defined by
   \be
   \hat \Lambda \equiv \langle
  \frac{  \hat \Delta \ {\bf v_f} \cdot   {\bf \nabla} T }
  { (|\Delta (\phi)|^2 + \gamma^2)^{3/2}} \rangle
  \label{Lambda}
  \ee
   which is finite only when $\hat \Delta$ is odd in $\hat p$.

   Substituting \eref{g1aK},\eref{g1aKns} and \eref{g1aKV} into \eref{s1aK} and
   observing \eref{g1aKnsav}, we see that $\sigma_{1a, \rm imp}^K$
   has the following form:
   \be
   \sigma_{1a, \rm imp}^K = \Gamma_u \left(
    \tilde X \tau_3 \hat \Lambda + i \tilde Y \hat \Lambda \right)
    \frac{ \gamma} { \mathcal{C}}
     \frac{\epsilon}{2 T^2} {\rm sech}^2 \frac{\epsilon}{2T}
   \label{XY}
   \ee
   with ${\rm Tr} \sigma_{1a, \rm imp}^K = 0$.
   For our particular form of $\hat \Delta$ and with
   the temperature gradient along $x$,
   we see that we would have
   \be
   \hat \Lambda = i \sigma_1 \tau_2 \Lambda_x  (dT/dx)
   \label{LLx}
   \ee
   \be
   \Lambda_x \equiv \langle \frac{  v_{f, x} \ \Delta_x }{D^3}
   \rangle
   \label{Lx}
   \ee

   If we only include $\hat g_{1a}^{K, \rm ns}$ in
   \eref{s1aK}, we would simply get, using \eref{g1aKnsav},
   $\tilde X \to X= ( \tilde \gamma^2 - {\rm cot}^2 \delta )$ and
   $\tilde Y \to Y= - 2 \tilde \gamma {\rm cot} \delta$.
   Including also $\hat g_{1a}^{K, V}$ of \eref{g1aKV}
   produces instead a pair of self-consistent equations
   for $\tilde X$ and $\tilde Y$, which can be
   written in a matrix form:
   \be
   \left(  \ba{cc}
   1 + \Gamma_u \frac{ \tilde \gamma^2 - {\rm cot}^2 \delta}
    {\mathcal{C}} \langle \frac{\Delta_x^2}{D^3} \rangle
    &
    2 \Gamma_u  \frac{ \tilde \gamma {\rm cot} \delta}{\mathcal{C}}
    \langle \frac{\Delta_x^2}{D^3} \rangle \\
    - 2 \Gamma_u  \frac{ \tilde \gamma {\rm cot} \delta}{\mathcal{C}}
    \langle \frac{\Delta_x^2}{D^3} \rangle
     & 1 + \Gamma_u \frac{ \tilde \gamma^2 - {\rm cot}^2 \delta}
    {\mathcal{C}} \langle \frac{\Delta_y^2}{D^3} \rangle
    \ea   \right)
    \left( \ba{c}
    \tilde X \\
    \tilde Y \ea \right)
    =     \left( \ba{c}
     X \\
     Y \ea \right)
     \label{XYmatrix}
    \ee
    with $X$ and $Y$ just defined above.
    Writing ${\rm Det} (>0)$ as the determinant of the matrix
    in \eref{XYmatrix}, evaluation of \eref{XYmatrix} simply gives us
    \be
    \tilde X = \left( \tilde \gamma^2 - {\rm cot}^2 \delta
    + \Gamma_u \langle \frac{\Delta_x^2}{D^3} \rangle \right)
    / {\rm Det}
    \label{tildeX}
    \ee
     and
    \be
    \tilde Y = - 2 \tilde \gamma {\rm cot} \delta / {\rm Det}
    \label{tildeY}
    \ee
    $\tilde Y$ is simply renormalized from $Y$ by the factor ${\rm Det}$.
    We shall see that $\tilde Y$ is responsible for the thermal Hall
    conductivity.

     The determinant can be simplified to be
       \be
       {\rm Det} = 1 + 2 \Gamma_u
       \langle \frac {\Delta_x^2}{D^3} \rangle
       \frac{ \tilde \gamma^2 - {\rm cot}^2 \delta}{\mathcal{C}}
       + \left( \Gamma_u
       \langle \frac {\Delta_x^2}{D^3} \rangle \right)^2
       \frac{1}{\mathcal{C}}
       \label{Det}
       \ee
      where we have used $\langle \frac {\Delta_x^2}{D^3} \rangle
      = \langle \frac {\Delta_y^2}{D^3} \rangle$ by tetragonal
      symmetry.  We shall see that ${\rm Det}$ is basically
      a numerical factor of order $1$.

      (In the above we have made use of the fact that $\hat \Delta$ 
      has a $\phi$ independent spin structure
      $\propto \sigma_1$ to simplify our calculations.  In the case
      of a more complicated $\phi$-dependent spin structure of the order parameter,
      if we are willing to ignore solving self-consistently the
      impurity self-energy, \eref{g1aKnsav}, \eref{Lambda}, \eref{XY}
      are still valid with the $\tilde X \to X$ and $\tilde Y \to Y$
      as mentioned above.  A more involved spin
      structure of the order parameter $\hat \Delta$ mainly
      modifies the value of the coefficients multiplying
      $\Gamma_u$ in \eref{XYmatrix}, though
      it can also generate more terms not within
      \eref{XY}. However, we expect that these complications
      would not lead to a significant change in the order of
      magnitude of the thermal Hall conductivity evaluated below).

    The energy current density along the $i$th direction is given by
    \be
    J^E_i = 2 N_f \int \frac{ d \phi}{ 2 \pi} v_{f, i}
    \int \frac{ d \epsilon} { 4 \pi i} \epsilon \
    \frac{1}{4} {\rm Tr} g^K_{1a}
    \ee
    where we have already dropped the contributions
    from the ``regular part" of $\hat g_1^K$ since
    it has no trace.  Using
    $\int d \epsilon \ \epsilon^2 \ {\rm sech}^2
    \frac{\epsilon}{2T} = \frac{4 \pi^2}{3}T^3$,
    the contribution to $J^E_i$ from
    $\hat g^{K, \rm ns}_{1a}$ is simply
    \be
    J^{E, \rm ns}_x = N_f \frac{\pi^2}{3} T \
     \langle v_{fx}^2 \frac{\gamma^2}{D^3} \rangle \ ( - \frac{dT}{dx} )
    \label{JEns}
    \ee
    which reproduces the result from \cite{Graf96}.
    The vertex correction contributions $\hat g_{1a}^{K, V}$
    generates the extra contributions, for temperature gradient
    only along x,
    \be
    J_x^{E, V} =  N_f \frac{\pi^2}{3} T \
     \Gamma_u \tilde X  \frac{\gamma^2}{\mathcal{C}}
     \Lambda_x^2 \ (\frac{dT}{dx} )
    \label{JExV}
    \ee
    and
   \be
    J_y^{E, V} =  N_f \frac{\pi^2}{3} T \
     \Gamma_u \tilde Y  \frac{\gamma^2}{\mathcal{C}}
     \Lambda_x \Lambda_y \ (\frac{dT}{dx} )
    \label{JEyV}
    \ee
    where $   \Lambda_y \equiv \langle \frac{  v_{f, y} \ \Delta_y }{D^3}
   \rangle$.
   For the last two contributions, we have used
   \bdm
   \frac{1}{4}{\rm Tr} g_{1a}^{K,V}
     = - \pi \Gamma_u \frac{i \gamma}{D^3} \
       \frac{1}{4}{\rm Tr}
      ( (\tilde X \hat \Lambda + i \tilde  Y \hat \tau_3 \hat \Lambda) \hat \Delta)
      \  \frac{\gamma}{\mathcal{C}}
      \frac{\epsilon}{2T^2}{\rm sech}^2 \frac{\epsilon}{2T}
      \edm
      in the $\epsilon \to 0$ limit
      and, for our specific
      choice of $\hat \Delta$,
       $ \frac{1}{4}{\rm Tr} (\hat \Lambda \hat \Delta) =
      - \Lambda_x \Delta_x \frac{dT}{dx}$ and
      $ \frac{1}{4}{\rm Tr} (\tau_3 \hat \Lambda \hat \Delta) =
      i \Lambda_x \Delta_y \frac{dT}{dx}$.

       We therefore get $J^E_x = K_{xx} (-dT/dx)$ and
       $J^E_y = K_{yx} (-dT/dx)$.  $K_{xx} = K_{xx}^{\rm ns} + K_{xx}^{V}$
        has two contributions
       respectively from \eref{JEns} and \eref{JExV}
       \be
       \frac{K_{xx}^{\rm ns}}{N_f \frac{\pi^2}{3} T} =
       \langle v_{fx}^2 \frac{\gamma^2}{D^3} \rangle
       \label{Kxxns}
       \ee
       \be
       \frac{K_{xx}^{V}}{N_f \frac{\pi^2}{3} T} =
       - \frac{\Gamma_u}{\rm Det}
       \left( (\tilde \gamma^2 - {\rm cot}^2 \delta) +
       \Gamma_u \langle \frac {\Delta_x^2}{D^3} \rangle \right)
       \frac{\gamma^2}{\mathcal{C}} \Lambda_x^2
       \label{KxxV}
       \ee
       but $K_{yx}$ arises from vertex corrections alone:
        \be
       \frac{K_{yx}^{V}}{N_f \frac{\pi^2}{3} T} =
       + \frac{\Gamma_u}{\rm Det}
       \left( 2 \tilde \gamma {\rm cot} \delta  \right)
       \frac{\gamma^2}{\mathcal{C}} \Lambda_x \Lambda_y
       \label{KyxV}
       \ee
       where we have used the results for $\tilde X$ and $\tilde Y$
       in \eref{tildeX} and \eref{tildeY}.
       \Eref{Kxxns} was obtained previously in \cite{Graf96}.
       Equations \eref{KxxV},\eref{KyxV} constitute the main results of
       this section.  We shall proceed to more specific
       forms of the gap ({\i.e.} $\eta_{x,y}$) below.

\subsection{Nodes}

\begin{figure}[htbp]
\begin{minipage}[b]{0.5\linewidth}
\centering
\includegraphics[width=\linewidth]{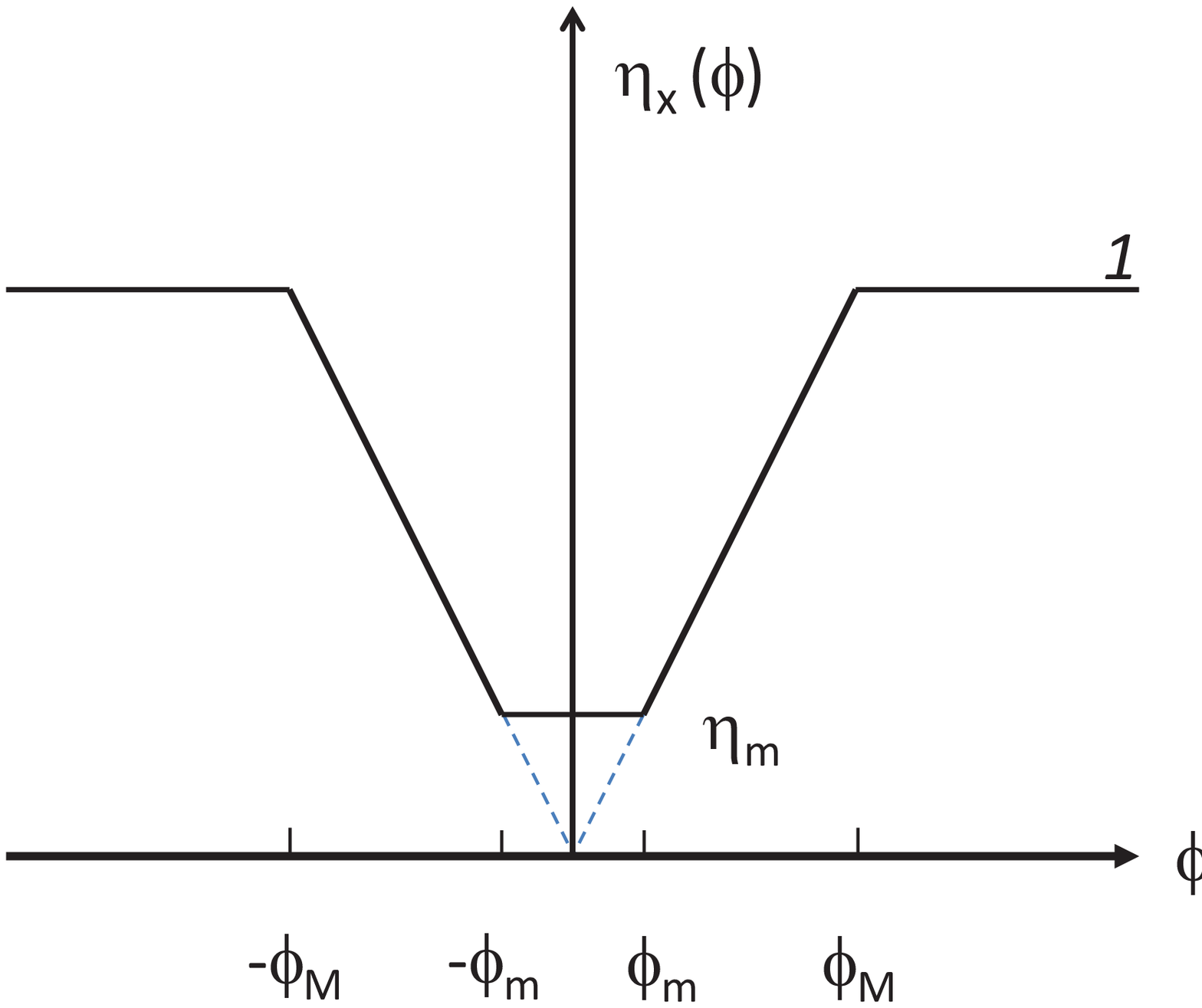}
\caption{The function $\eta_x(\phi)$ near $\phi = 0$}
\label{Fig1}
\end{minipage}%
\hspace{0.1cm}
\begin{minipage}[b]{0.5\linewidth}
\centering
\includegraphics[width=\linewidth]{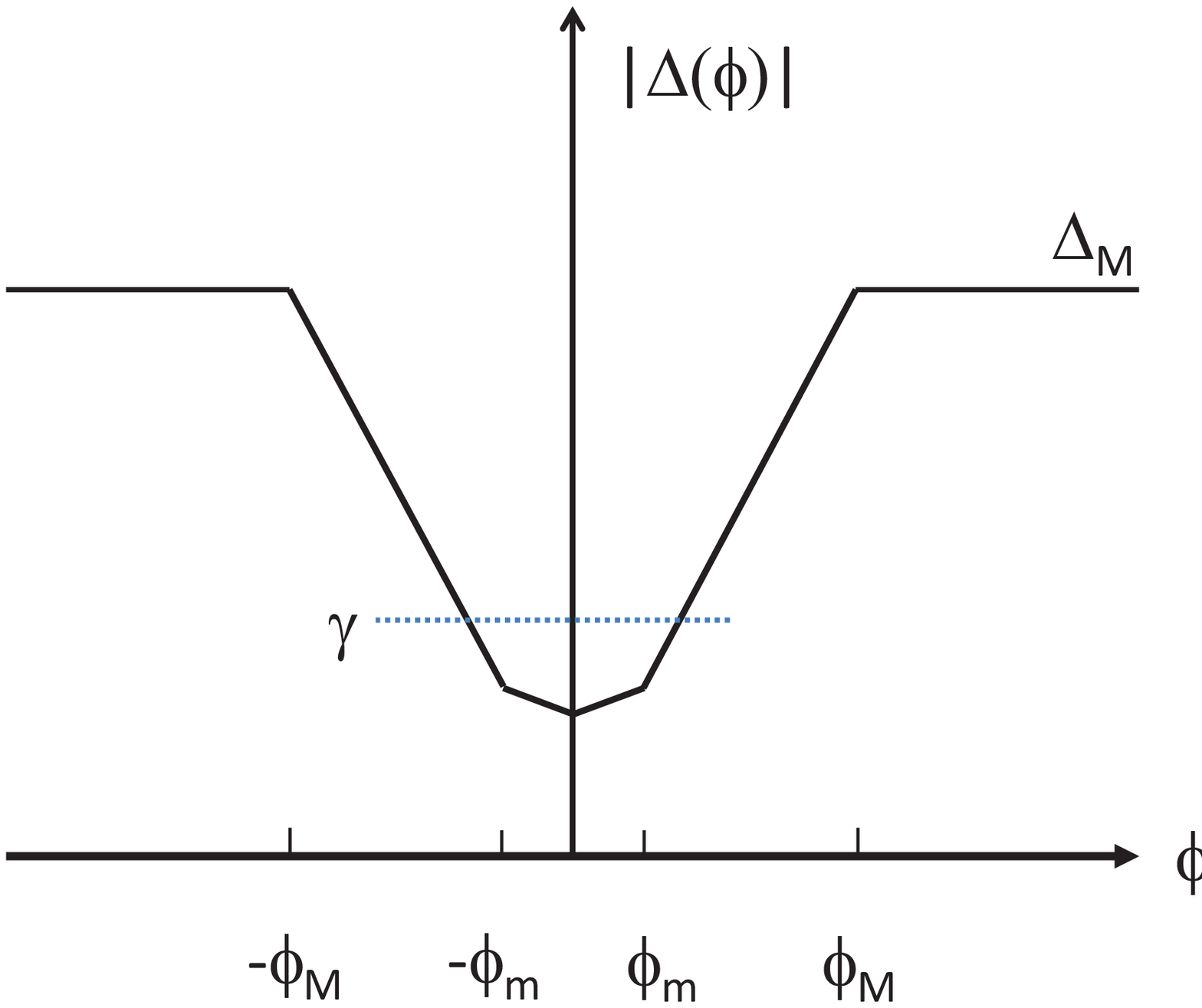}
\caption{The magnitude of the gap $|\Delta(\phi)|$ near $\phi = 0$.}
\label{Fig2}
\end{minipage}
\end{figure}

For definiteness,
we now consider specific models of the gap, {\i.e.},
special form of the functions $\eta_{x,y}(\phi)$.
We would like to have models which would exhibit
universal thermal conductivity at low temperatures,
at least for $\gamma$ within certain range.
We first consider the particular case where these $\eta_{x,y}$ factors
do not introduce extra sign changes in the order parameter.
(This model is motivated by the proposal that
the gap is suppressed for $\hat {\bf p} \sim \hat x$
and $\hat y$ for the $\gamma$ band since these points
are close to the Brillouin zone boundary, e.g., \cite{Nomura02,Yanase03}.
For generalizations of this model, see \ref{sec:AppB}).
We shall take, for $\phi$ within the first quadrant
\begin{equation}
\label{etax}
\eta_x (\phi)
=\cases{\eta_m = \phi_m/\phi_M &  $ 0 < \phi < \phi_m $\\
 \phi/\phi_M &  $ \phi_m < \phi < \phi_M$ \\
 1&  $\phi_M < \phi < \pi/2 $\\}
\end{equation}
where thus $\eta_x$ takes its minimum value $\eta_m = \phi_m/\phi_M$
for small $\phi$,
with $\eta_x$ linear in $\phi$ up to some maximal
value $\phi_M$ (which is of order $1$ but less than $1$),
beyond which $\Delta_x$ is not modified
from the isotropic model where $|\Delta(\phi)| = \Delta_M$
(see Fig \ref{Fig1}).
For the other quadrants, we define $\eta_x$ such that
$\eta_x$ is symmetric under $p_y \to - p_y$
and $p_x \to - p_x$
(hence $\eta_x$ invariant under
$\phi \to - \phi$ and $\phi \to \pi - \phi$.
To maintain the transformation properties of
$\Delta_{x,y}(\phi)$, we choose
$\eta_y(\phi+\pi/2) = \eta_x(\phi)$).
With this $\eta_x$ and $\eta_y$, we have produced a near-node
near $\phi = 0$ (and a true node at $\phi=0$ if $\eta_m$ and $\phi_m$
approach zero, and similarly for $\phi$ near $\pi/2$ etc).
If $\phi_M$ is small, we have
approximately
\begin{equation}
\label{D2}
|\Delta (\phi)|^2
\approx \cases{|\Delta_M|^2 (\eta_m^2 + \phi^2) &  $ 0 < \phi < \phi_m $\\
 |\Delta_M|^2 \left( \frac{1}{\phi_M^2} + 1 \right) \phi^2&
  $ \phi_m < \phi < \phi_M$ \\
 |\Delta_M|^2 &  $\phi_M < \phi < \pi/2 $\\}
\end{equation}
Furthermore, for $\phi_m < \phi < \phi_M$,
$\left( \frac{1}{\phi_M^2} + 1 \right) \phi^2 \approx
\left(\phi/\phi_M\right)^2$, an approximation that
we shall also take for simplicity.
The form for $|\Delta(\phi)|$ is shown schematically in
Fig. \ref{Fig2}.  For reasons that would be clear immediately,
we shall require that $\gamma$ is large compared the
energy gaps in the $|\phi| < \phi_m$ region, thus
$\Delta_M (\phi_m / \phi_M) \ll \gamma$ (and hence also
$\Delta_M \phi_m \ll \gamma$), but $\gamma \ll \Delta_M$,
as indicated also schematically in Fig \ref{Fig2}.

We evaluate the various angular averages that appear in
our expressions \eref{Kxxns},\eref{KxxV} and \eref{KyxV}
for thermo-conductivities.  Consider first the quantity
in \eref{Kxxns}.  We can rewrite it as
\bdm
\frac{2}{\pi} v_f^2 \int_0^{\pi/2} d \phi \frac{ \gamma^2 {\rm cos}^2 \phi}
{\left\{|\Delta(\phi)|^2 + \gamma^2 \right\}^{3/2}}
\edm
and separate the integration into three regions as in \eref{etax}.
For the integral in the first region, we need (we left out the common
factor $ (2/\pi) v_f^2$ for simplicity)
\bdm
\int_0^{\phi_m} d \phi \frac{ \gamma^2 {\rm cos}^2 \phi}
{\left\{|\Delta_M|^2 ( \eta_m^2 + \phi^2) + \gamma^2 \right\}^{3/2} }
\edm
If the inequalities concerning $\gamma$ mentioned below
\eref{D2} holds,
we can ignore all terms other than $\gamma$ in the denominator.
This contribution is therefore of order $\phi_m/\gamma$.
The integral for the second contribution can be evaluated.
If the above stated inequalities are also satisfied,
was stated below \eref{etax}, we have
\be
\int_{\phi_m}^{\phi_M} d \phi \frac{ \gamma^2 {\rm cos}^2 \phi}
{\left\{|\Delta_M|^2 ( \frac{\phi}{\phi_M^2})^2 + \gamma^2 \right\}^{3/2} }
\approx \frac{\phi_M}{\Delta_M}
\ee
The third contribution is of order $\gamma^2/\Delta_M^3$.
Again using the inequalities stated below \eref{etax},
the second contribution dominates and
we have the approximation
\be
 \langle v_{fx}^2 \frac{\gamma^2}{D^3} \rangle \approx
 v_f^2 \ \frac{2 \phi_M}{\pi \Delta_M}
 \label{uap}
\ee
We see that we have
\be
\frac{K_{xx}^{ns} }{ N_f v_f^2 \frac{\pi^2}{3} T}
\approx \frac{ 2 \phi_M}{ \pi \Delta_M}  \ ,
\label{univ}
\ee
the universal value
in \cite{Graf96} but in slightly different notations.
Thus this model of the gap is consistent
with the experimental observations \cite{Suzuki02,Suderow98},
even though the (near) nodes here are not produced by
the sign change of the order parameter (as in the
case of e.g. $d_{x^2 - y^2}$ in \cite{Graf96}).

Now we turn to $K_{yx} = K_{yx}^{V}$, our main interest in this paper.
(For $K_{xx}^V$, see \ref{sec:AppA}.)
Eliminating $\Gamma_u$
 in favor of $\tilde \gamma$ and ${\rm cot} \delta$,
 we obtain
        \be
       \frac{K_{yx}^{V}}{N_f \frac{\pi^2}{3} T} =
       + \frac{\gamma^3}{\rm Det}
       \left(
       \frac{2 {\rm cot} \delta}{\tilde \gamma^2 +  {\rm cot}^2 \delta}\right)
        \Lambda_x \Lambda_y
       \label{KyxV2}
       \ee
with
 \be
 {\rm Det} = 1 + 2 \ \frac{\gamma}{\tilde \gamma} \
                          \frac{ {\tilde \gamma^2} - {\rm cot}^2 \delta}
                               { {\tilde \gamma^2} + {\rm cot}^2 \delta}
                            \langle \frac{\Delta_x^2}{D^3} \rangle
               + \frac{\gamma^2}{\tilde \gamma^2}
                           \langle \frac{\Delta_x^2}{D^3} \rangle^2
 \label{Det2}
 \ee

The required angular averages can be obtained in a similar manner as just
described for \eref{Kxxns}.  We have
\be
\Lambda_x \approx \frac{2 v_f}{\pi} \frac{\phi_M}{\gamma \Delta_M}
\label{Lxap}
\ee
with the correction term of order $1/\Delta_M^2$,
$\Lambda_x = \Lambda_y$ by symmetry, and
\be
\langle \frac{\Delta_x^2}{D^3} \rangle \approx \frac{2 \phi_M}{\pi \Delta_M}
 \ln \frac{ 2 \Delta_M}{\gamma}
 + \frac{1 - \frac{2 \phi_M}{\pi/2}}{2 \Delta_M}
 \label{Dx2ap}
 \ee

 \be
 \tilde \gamma \approx \frac{4 \gamma}{\pi}
 \left( \frac{\phi_M}{\Delta_M} \ln \frac{ 2 \Delta_M}{\gamma}
   + \frac{\pi/4 - \phi_M}{\Delta_M} \right)
 \label{tgap}
 \ee
 In the above two expressions, the second terms are
 only logarithmically smaller than the first.

 With the above approximations, one can check that
 the quantity ${\rm Det}$ defined in \eref{Det}
 is only weakly dependent on the phase
 shift $\delta$.    Keeping only the dominant terms in
 \eref{Dx2ap} and \eref{tgap} and using \eref{tgamma}
 we see that ${\rm Det} \approx 9/4$ near resonance
 and ${\rm Det} \approx 1/4$ in the Born limit.
 Suppressing this factor, we get
 \be
 K_{yx}/K_{xx}^{ns} \approx
 \frac{2 \phi_M}{\pi}
 \frac{\gamma}{\Delta_M} \
       \frac{2 {\rm cot} \delta}
       { {\rm cot}^2 \delta + (\frac{ 4 \phi_M}{\pi}
       \frac{\gamma}{\Delta_M} \ln \frac{2 \Delta_M}{\gamma} )^2}
 \label{Kyxap2}
 \ee
  The last fraction is small in both the Born and resonance limit,
  a result already known in \cite{Arfi}.  If however
  $|{\rm cot} \delta| \sim \tilde \gamma $
  ($\approx  \frac{ 4 \phi_M}{\pi}
       \frac{\gamma}{\Delta_M} \ln \frac{2 \Delta_M}{\gamma} $),
  we have instead
  $ |K^{yx}/K_{xx}^{ns}| \approx 1/ (2 \ln \frac{2 \Delta_M}{\gamma})$,
  hence the off-diagonal term is only logarithmically smaller
  than the universal diagonal term.

  The value of $\gamma$ depends on the impurity concentration etc.
  As a reference, \cite{Suzuki02} estimates that for their samples
  measured there, $\gamma$ ranges from $~0.2K$
  to $~ 0.6K$ (while still observing thermal conductance close
  to the universal value).   Estimating $2 \Delta_M$ by say $ 5 T_c$,
   this logarithmic factor is only roughly $0.13$ and $0.2$ for these $\gamma$'s.

  Let us further examine the magnitude of $K_{yx}$.  Using the
  expression for $K_{xx}^{ns}$, and say ${\rm cot} \delta$ is
  at the above optimal value, we have the following estimate:
  \be
  \frac{K_{yx}}{T} \sim \frac{K_{xx}^{ns}}{2 T \ln \frac{2 \Delta_M}{\gamma}}
  \sim \frac{\pi^2}{6}
  k_B^2 \frac{ N_f v_f^2 \ \frac{ 2 \phi_M}{\pi \Delta_M}}
    {\ln \frac{2 \Delta_M}{\gamma}}
    \label{com}
    \ee
    In the last expression, we have restored the Boltzmann constant
    $k_B$.
    Since $N_f v_f^2 / \Delta_M \sim k_f v_f / \Delta_M
    \sim E_f / T_c$ where $E_f$ is the Fermi temperature
    and $T_c$ the superconducting critical temperature,
    we see that the numerator in \eref{com} is large.
    (For SrRu$_2$O$_4$, this ratio is probably of order $10^2$).

\begin{figure}[htbp]
\centering
\includegraphics[width=0.8\linewidth]{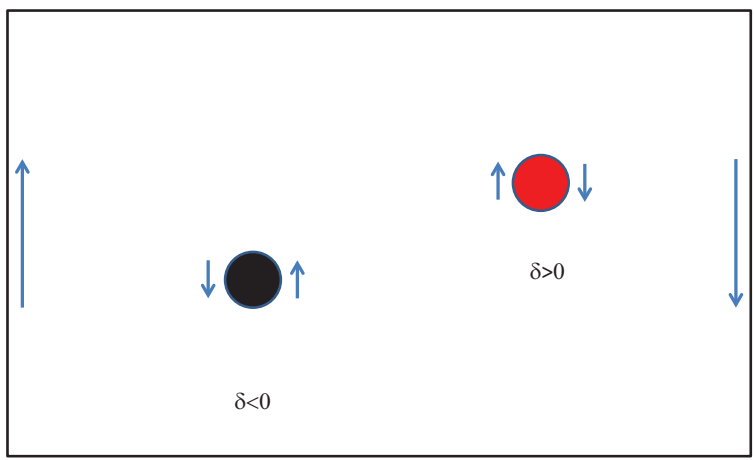}
\caption{Edge states of a chiral $p_x + i p_y$ superconductor, with
$x$ axis horizontal and $z$ axis pointing out of the paper.  Bound
states at the left edge of the sample disperse upwards (along $+\hat y$)
and vice versa at the right edge
(see, e.g. \cite{Stone}, note particle current on the left would
rather be expected to be downwards along $-\hat y$).  Also shown are the
edge states near
a region with repulsive (labeled by $\delta < 0$)
and attractive (labeled by $\delta > 0$) potential inside the sample. }
\label{Fig3}
\end{figure}

     Hence, unless the phase shift $\delta$ is very close to
    multiple of $\pi/2$, our computed thermal Hall conductivity
    is large compared with $k_B^2 T$.  Recently, in the context
    of topological superconductivity, the contribution of
    edge states to the thermal Hall conductivity is also much
    discussed \cite{Scaffidi15} (see also e.g.,
    \cite{Read,Senthil99,Nomura12,Sumiyoshi}).
    Assuming the bulk of a chiral superconductor
    is gapped, the topological edge states become the
    sole carriers for the heat current.      See Fig \ref{Fig3}.
    Consider a sample where say the left is hotter than the right,
    a net thermal current is then generated in the direction
    perpendicular to the temperature gradient
    (upwards along $\hat y$ in Fig \ref{Fig3})
   This thermal Hall conductivity
    is of order $k_B^2 T$.   While this claim is probably correct,
    we note that the observation of this thermal Hall conductivity
    is very restrictive. If an impurity band forms
    (which seems the case for all Sr$_2$RuO$_4$ samples studied so far), then
    the scenario of the present paper applies,  giving rise
    to a much larger thermal Hall conductivity, though
    the thermal Hall angle would be much smaller since now $K_{xx}$ is finite.
    (see also \ref{sec:AppB}.)

  Let us now discuss the sign of this thermal Hall conductivity.
  Our results shows that the sign of $K_{yx}$ is the same
  as $  {\rm cot} \delta$.  (e.g.,
  $K_{yx} > 0$ ($K_{xy} = - K_{yx} < 0$) if $0 < \delta < \pi/2$).
  It seems that
  this {\em sign} of the thermal Hall conductivity can be
  understood in the same manner as the edge state contributions
  \cite{Read,Senthil99} mentioned above.
  Consider extended impurities (rather than point impurities in
  our calculations) within the sample.
  A repulsive potential
 inside the sample has edge states propagating down ($-\hat y$)
 on its left but up on its right.    The opposite situation
 applies if the potential is attractive.
   If the sample is hotter on
the left and cooler on the right ($ - (dT/dx) > 0$),
the edge states near the attractive (repulsive) potential wells would
generates a net heat current along $+\hat y$ ($-\hat y$), if we have many
such potential wells (barriers) such that these bounds states
overlap so that the heat can propagate through the sample.
Now for point impurities with attractive (repulsive) potentials,
$\delta > (<) 0$ \cite{Sakurai}.
(More precisely our argument is restricted to
$0 < \delta < \pi/2$ for attractive potentials and
$ - \pi/2 < \delta < 0$ for repulsive potentials.  Note however
if say $ \pi/2 < \delta < \pi$,
scattering behavior of particles by the attractive potential is
as if the scattering were from a repulsive potential with
$ -\pi/2 < \delta < 0$.)
 We thus see that the sign of the thermal Hall
conductivities we obtained are of the same sign as what
we have if the potentials are extended.

    Arfi et al discussed the state $\hat z (p_x + i p_y)$.
    In their plots, they give $K_{xy} > 0$ with phase shifts
  $ 0 < \delta < \pi/2$, thus with a sign opposite to the present
  results.  The author has not yet been able to resolve this discrepancy.

\section{Summary and Discussion}\label{sec:sum}

In this paper, we have evaluated the $T \to 0$ thermal Hall conductivity
for a chiral superconductor where there are momentum directions
with strong gap suppression, assuming simple isotropic
impurities.   We find that the value of this thermal Hall conductivity can be
a significant fraction of the diagonal universal component.
Detection of this thermal Hall conductivity in Sr$_2$RuO$_4$ would be a strong
proof that the superconductor possesses a chiral p-wave order parameter.
This experimental measurement can however be quite challenging,
as it would require a sample with a single domain order parameter.
At this moment, there are quite some uncertainties about
the domain sizes \cite{VH06,Kallin09,Hicks10} in this superconductor,
but measurement techniques applicable to small system sizes
would definitely help here.

We have only considered isotropic impurities, and have ignored
possible anisotropic scattering by the impurities,
spin-orbit couplings both for the bulk and the impurity scattering.
It is entirely feasible that these would
also generate a finite thermal Hall conductivity, provided
the superconducting order parameter breaks time-reversal
and inversion symmetry.   The investigation of these possibilities
however must be left for the future.

\ack
This research is supported by the Ministry of Science and Technology
of Taiwan under grant numbers NSC 101-2112-M-001-021-MY3 and
MOST 104-2112-M-001-006-MY3.

\appendix

\section{Vertex Corrections $K_{xx}^{V}$}\label{sec:AppA}
\setcounter{section}{1}

Though not directly related to the central theme of this paper,
we here investigate further the vertex corrections
$K_{xx}^{V}$ to the diagonal terms of the thermal conductance,
as we do not find much calculations of this in the literature, with
the exception of \cite{Durst} for the $d_{x^2-y^2}$ superconductor
and the specific model of internodal scattering.
The needed expression is already given in \eref{KxxV}.
For simplicity we shall confine ourselves only
to the Born and resonance limits.
Again eliminating $\Gamma_u$ in favor of $\tilde \gamma$ and ${\rm cot} \delta$,
and setting $\delta \to 0$,
we get (using ${\rm Det} \to 1/4$, note that in this limit,
the second and third terms in \eref{tildeX} dominates)
       \be
       \frac{K_{xx}^{V}}{N_f \frac{\pi^2}{3} T} \approx
        \frac{ 4 \gamma^3}{\tilde \gamma^2}
       \left( (\tilde \gamma - \gamma
        \langle \frac {\Delta_x^2}{D^3} \rangle \right) \Lambda_x^2
       \label{KxxVB}
       \ee
  We see that this is positive definite, that is, the
  thermal conductivity becomes larger than the universal value
  when this vertex correction is included.
  Using the approximate formulas \eref{Dx2ap} and \eref{tgap}
  we get
  \be
  K_{xx}^{V}/K_{xx}^{ns} \approx 1 / \ln (2 \Delta_M/\gamma)
  \label{KxxVB2}
  \ee
  On the other hand, in the resonance limit, we get instead
         \be
       \frac{K_{xx}^{V}}{N_f \frac{\pi^2}{3} T} \approx
        - \frac{4 \gamma^3}{9 \tilde \gamma^2}
       \left( \tilde \gamma +  \gamma
        \langle \frac {\Delta_x^2}{D^3} \rangle \right) \Lambda_x^2
       \label{KxxVR}
       \ee
   This is then negative.   Using the approximate values
   discussed before we get
   \be
   \frac{K_{xx}^V}{K_{xx}^{ns}} \approx  - 1 / 3  \ln (2 \Delta_M/\gamma)
   \label{KxxVR2}
   \ee
    The above mentioned sign change in vertex corrections do not
    seem to have been noted before in the literature.
     Durst and Lee \cite{Durst}, who
    did investigate the vertex corrections for a model of
    internodal scattering in $d_{x^2 - y^2}$ superconductor,
    did not mention such possibility of sign change.
    It is
    difficult to compare our calculations with \cite{Durst}
    since the model are quite different.
    However, we do note here
    that the coefficients defined in their (3.17a) and (3.17b)
    entering their expression for the vertex correction (4.25)
    for thermal conductivity also consist of two terms appearing
    as differences, so a sign change may not be out of the question.

\section{Other models of the gap}\label{sec:AppB}
\renewcommand{\theequation}{B.\arabic{equation}}
\setcounter{equation}{0}
\renewcommand{\thesection}{B}
\setcounter{subsection}{0}

\subsection{more general forms of $\eta_{x,y}$}

For definiteness, in the text we have introduced a specific
model for the functions $\eta_{x,y}$ to introduce near-line
nodes in the gap.   We shall now argue that many of our
results are basically unchanged for other forms of $\eta_{x,y}$
so long as we still have near-line nodes, provide some
rather weak conditions remain satisfied.  Thus the results
given in the text are quite general.

For example, our results are directly applicable to the case
where $\eta_x$ in Fig \ref{Fig1} in the region $\phi_m < \phi < \phi_M$
is not directly proportional to $\phi$ but rather have
a form say $ (1 - \phi_M/\phi'_M) + \phi/ \phi'_M$,
which is still linear in $\phi$ but extrapolate rather
to a finite value at $\phi=0$.   If the angular averages
in \eref{uap},\eref{Lxap},\eref{Dx2ap},\eref{tgap} etc are still
dominated by the contributions where the gap is linear in $\phi$,
these equations are then essentially unchanged and our
estimate for the thermal Hall conductivities remain valid.
With the same reasoning, $\Delta_x$ and $\Delta_y$ can have
additional sign changes as functions of $\phi$ so long
as the above mentioned integrals are dominated by the linear
regions, in particular $\Delta(\phi) \equiv \Delta_x + i \Delta_y$ as a function
of $\phi$ can have winding number larger than $1$, so long as
the angular average \eref{Lxap}, which involves
$\sim \langle \frac{ v_x \Delta_x}{(\tilde \gamma^2 + |\Delta|^2)^{3/2}} \rangle$
 is non-zero.
(For example, if $\Delta_x = {\rm cos} (3 \phi)$ and
$\Delta_y = {\rm sin} (3 \phi)$, then
the vertex corrections would vanish for isotropic Fermi surface
and so $K_{yx}$ would become zero.  However, for more
general forms of $\Delta(\phi)$ so that $d \Delta(\phi)/d \phi$
or the magnitude $|\Delta(\phi)|$ is
not a constant in $\phi$, \eref{Lxap} is in general finite
even if $\Delta(\phi)$ still has winding number $3$ or
other odd numbers).

Our calculations are valid also for the near nodes are located
along $\hat k = \pm \hat x'$ and $\pm \hat y'$ where $\hat x'
= (\hat x + \hat y)/\sqrt{2}$ and $\hat y' = (\hat y - \hat x)/\sqrt{2}$.
This is because we are only evaluating linear response and
we could have use coordinates $\hat x'$ and $\hat y'$ for
our calculations in text.  Tetragonal symmetry implies
that $K_{x'x'} = K_{xx}$ and $K_{y'x'} = K_{yx}$ etc.
Hence our estimates for the thermal conductance should
also be applicable for the models in e.g.
\cite{Wang13,Scaffidi15}.

\subsection{the fully gapped case}

Though not the main concern of the present paper, we discuss
here also the  $T \to 0$ thermal conductance for a two-dimensional
chiral superconductor
with full isotropic gap, with $\Delta_x = \Delta_M \cos \phi$
and $\Delta_y = \Delta_M \sin \phi$.  The required angular
averages are trivial in the $\gamma \ll \Delta_M$ limit.
We have
\be
 \langle v_{fx}^2 \frac{\gamma^2}{D^3} \rangle \approx
 v_f^2 \ \frac{\gamma^2}{2 \Delta_M^3}
 \label{Buap}
\ee
Hence
\be
\frac{K_{xx}^{ns} }{ N_f v_f^2 \frac{\pi^2}{3} T}
\approx \frac{ \gamma^2} {  2 \Delta_M^3  } ,
\label{BKxx}
\ee
which is no longer universal, and is smaller
than the text for the case of a nodal or near-nodal
superconductor by a factor of $~ (\gamma/\Delta_M)^2$.
  Furthermore,
\be
\tilde \gamma \approx \gamma / \Delta_M \ ,
\ee
\be
\Lambda_x = \Lambda_y \approx \frac{v_f}{2 \Delta_M^2}
\label{BLxap}
\ee
and
\be
\langle \frac{\Delta_x^2}{D^3} \rangle \approx \frac{1}{2 \Delta_M}
 \label{BDx2ap}
 \ee
  We can verify that ${\rm Det}$ defined in \eref{Det} is
  again $~1/4$ and $~9/4$ in the Born and resonant limit respectively.
  Suppressing again this numerical constant,  we obtain
  \be
 \frac{K_{yx}}{N_f v_f^2 \frac{\pi^2}{3} T} \approx
       \frac{2 {\rm cot} \delta}
       { {\rm cot}^2 \delta +
       \frac{\gamma^2}{\Delta_M^2} }
       \frac{\gamma^3}{\Delta_M^4}
 \ee
 If $\rm cot \delta \sim \gamma/\Delta_M$, then it turns
 out that $K^{yx} \sim K_{xx}^{ns}$.
 Similarly, one can verify that $K_{xx}^{V}$ becomes
 the same order as $K_{xx}^{ns}$.

\end{document}